\documentclass[12pt]{iopart}
\usepackage{bm,color,bbm}

\usepackage{ulem} %for strike-through font, e.g \sout{Bush} Obama is the president. 
\expandafter\let\csname equation*\endcsname\relax
\expandafter\let\csname endequation*\endcsname\relax   

\usepackage{hyperref, mathtools,graphicx,amsmath,cases}

\newcommand{\beq}{\begin{equation}}
	\newcommand{\eeq}{\end{equation}}
\newcommand{\bqa}{\begin{eqnarray}}
	\newcommand{\eqa}{\end{eqnarray}}
\newcommand{\nn}{\nonumber}

\newcommand{\smallfrac}[2]{\mbox{$\frac{#1}{#2}$}}

\newcommand{\half}{\smallfrac{1}{2}}

\newcommand{\blk}{\color{black}}

\definecolor{maroon}{rgb}{0.7,0,0}
\definecolor{ngreen}{rgb}{0.3,0.7,0.3}

\definecolor{golden}{rgb}{0.8,0.6,0.1}

\usepackage{iopams}  

\begin{document}
	
	\title{Comment on  ``Vindication of entanglement-based witnesses of non-classicality in
		hybrid systems''}
	%\title[Comment on  ``Vindication of entanglement-based witnesses of non-classicality \dots'']{Comment on  ``Vindication of entanglement-based witnesses of non-classicality in
	%	hybrid systems''}

	\author{Michael J. W. Hall$^1$ and Marcel Reginatto$^{2}$}
	\address{$^1$ Department of Theoretical Physics, Research School of Physics, Australian National University, Canberra ACT 0200, Australia}
	\address{$^2$ Physikalisch-Technische Bundesanstalt, Bundesallee 100, 38116 Braunschweig, Germany}
	
	\vspace{10pt}

	\begin{abstract}  Models of quantum-classical interactions fall into two classes: those which allow the generation of quantum entanglement via a classical mediator (such as gravity), and those which do not. Marconato and Marletto have recently sought to distinguish between these classes by claiming that known members of the first class (based on the configuration-ensemble formalism introduced by us) fail to model the mediator as a `classical' system, and are nonlocal. We explicitly show that this claim is incorrect, and expose a large number of errors and misconceptions in their reasoning. We also point to a very simple and transparent example of the generation of entanglement between two qubits via a classical bit.  It follows that there are models permitting the generation of entanglement via quantum-classical interactions that lie outside the remit of the theorem cited by Marconato and Marletto. We clarify the reasons for the limited applicability of various no-go theorems for entanglement generation.  
	\end{abstract}
	
	%
	% Uncomment for keywords
	%\vspace{2pc}
	%\noindent{\it Keywords}: Bell nonlocality, Horodecki criterion, CHSH inequality, POVMs\\
	%
	% Uncomment for Submitted to journal title message
%	\submitto{\JPA}
	%
	% Uncomment if a separate title page is required
	\maketitle
	% 
	% For two-column output uncomment the next line and choose [10pt] rather than [12pt] in the \documentclass declaration
	%\ioptwocol
	%
	
	\section{Introduction}
	\label{sec:intro}
	
	The generation of entanglement between two quantum systems, via local interactions with a classical mediator, is incompatible with several models of quantum-classical interaction~\cite{Kafri,Bose, MVprl,HR2017, MV2020}. It follows in particular that if a gravitational interaction was observed to  generate quantum entanglement between two masses, then this would rule out a number of models of classical gravity.  The likelihood of being able to make such an observation within the next decade or so~\cite{Bose} has attracted considerable interest (see, e.g.,~\cite{CR2019,KTPP2020,Howl21,CMT2021} and references therein).
	
	Nevertheless, there are models of quantum-classical interaction, proposed by us, that {\it can} generate entanglement via classical mediation, and thus cannot be ruled out by the observation of such entanglement~\cite{HR2017}. In a recent paper Marconato and Marletto have claimed that this class of models can instead be ruled out {\it a priori}, on the grounds that they are nonlocal and do not provide a classical description of the mediator~\cite{MM}. In section~\ref{sec:errors} we show this claim is incorrect because it is based on a large number of errors and misconceptions, which we explicitly point out. 
	
	Macconato and Marletto further claim, incorrectly, that a theorem in Ref.~\cite{MV2020} forbids the possibility of entanglement generation via a classical mediator for {\it any} model that satisfies two particular principles~\cite{MM}. As this more general claim is the main motivation for their paper (they wish to `vindicate' the theorem in the light of our models), we address it in section~\ref{sec:assump}. In particular, we point out our models are perfectly compatible with this theorem, because it relies on many additional assumptions that need not be satisfied by physical models (and which are not satisfied in our models). We further illustrate this reliance on additional assumptions by a very simple example of the generation of entanglement between two qubits via a classical bit, where this example satisfies both of the principles referred to by Marconato and Marletto. Note that section~\ref{sec:assump} can be read independently of section~\ref{sec:errors}, and hence the interested reader may directly proceed to it if they so wish.

	\section{Errors and misconceptions}
	\label{sec:errors}
	
	\subsection{The model criticised by Marconato and Marletto}
	\label{sec:model}
	
	We will confine our discussion here to the particular model in~\cite{HR2017} that Marconato and Marletto attempt to criticise in~\cite{MM} (other models, including the generation of quantum field entanglement via classical general relativity, have also been given~\cite{HR2017,HRgrav}). We first briefly collect some necessary formulas (with some minor notational changes to~\cite{HR2017} and~\cite{MM} that we hope will add clarity to the discussion). 
	
	The model describes quantum-classical interactions via the formalism of ensembles in configuration space~\cite{HR2005,PRA2008,HRbook}, for the case of two quantum particles $Q_1$ and $Q_2$ and a classical particle $C$ each moving in one dimension. Such an ensemble is described by a probability density $P(q_1,q_2,x)$ and a canonically conjugate quantity $S(q_1,q_2,x)$ on the configuration space,  where $q_1, q_2$ and $x$ label the position coordinates of particles $Q_1, Q_2$ and $C$.  Observables are represented by suitable functionals of $P$ and $S$, with a numerical value equal to the ensemble average of a corresponding measurement of the observable. In particular, each function $f(x,k)$ on the classical phase space of $C$ (where $x$ and $k$ are classical position and momentum coordinates)  and each Hermitian operator $\hat M$ on the Hilbert space of $Q_1$ and $Q_2$ are represented by the respective configuration-ensemble observables 
	\begin{align} \label{cf}
	\langle f\rangle &= C_f[P,S]:=\int dq_1dq_2dx\, P(q_1,q_2,x) f(x,\partial_x  S(q_1,q_2,x)), \\
    \label{qm}
	\langle \hat M\rangle &= Q_{\hat M}[P,S]:=\int dx\, P(x) \int dq_1dq_2\, \psi^*(q_1,q_2|x)\hat M\psi(q_1,q_2|x), 
	\end{align}
	where $P(x):=\int dq_1dq_2 P(q_1,q_2,x)$ is the marginal position density for the classical particle $C$, and $\psi(q_1,q_2|x):=\sqrt{P(q_1,q_2,x)/P(x)} e^{iS(q_1,q_2,x)/\hbar}$ defines a wave function on the Hilbert space of $Q_1$ and $Q_2$ for each value of $x$.  It follows immediately from~(\ref{cf}) and~(\ref{qm}) that the classical and quantum statistics are equivalently described by the classical phase space function $\rho_{C}(x,k)$ and  quantum density operator $\hat\rho_{Q_1Q_2}$ defined by
	\beq \label{equiv}
	\rho_{C}(x,k):= \int dq_1dq_2\,P(q_1,q_2,x)\delta(k-\partial_x  S(q_1,q_2,x)),~~ \hat\rho_{Q_1Q_2}:=\int dx\,  P(x)  |\psi_x\rangle\langle\psi_x|,
	\eeq
 where the ket $|\psi_x\rangle$ corresponds to the wave function $\psi(q_1,q_2|x)$.
	
	The Poisson bracket of any two observables $V[P,S]$ and $W[P,S]$ is defined by
	\beq
	\left\{V,W\right\} := \int dq_1dq_2dx\,\left(\frac{\delta V}{\delta P} \frac{\delta W}{\delta S} - \frac{\delta W}{\delta P} \frac{\delta V}{\delta S}\right),
	\eeq
	where $\delta/\delta P$ and $\delta/\delta S$ denote functional derivatives with respect to the conjugate quantities $P$ and $S$, and one can show that~\cite{HRbook}
	\beq \label{brackets}
	\{C_f,C_g\} = C_{\left(\frac{\partial f}{\partial x}\frac{\partial g}{\partial k}-\frac{\partial g}{\partial x}\frac{\partial f}{\partial k}\right)},\qquad \{Q_{\hat M},Q_{\hat N}\} = Q_{i[\hat M,\hat N]/\hbar} \, .
	\eeq
	Thus, the bracket algebra of classical observables  $C_f[P,S]$  is isomorphic to the usual Poisson bracket  algebra of functions $f(x,k)$  on classical phase space, while that of quantum observables  $Q_{\hat M}[P,S]$ is isomorphic to the usual commutator algebra of operators  $\hat M$  on Hilbert space. Further, the evolution of the ensemble is determined by a Hamiltonian functional,  the \textit{ensemble Hamiltonian}  $H[P,S]$, via the Hamilton equations
	\beq \label{motion}
	\frac{\partial P}{\partial t} =  \frac{\delta H}{\delta S},
	\qquad \qquad
	\frac{\partial S}{\partial t} =  -\frac{\delta H}{\delta P}, 
	\eeq
	implying that an arbitrary observable $V[P,S]$ (and hence its average value) evolves as
	\beq \label{dvdt}
	\frac{dV[P,S]}{dt} = \{ H, V\} .
	\eeq
	
Our model of entanglement generation in section~4.2 of~\cite{HR2017} corresponds to a quantum-classical interaction described by the ensemble Hamiltonian
	\beq \label{ham}
	H[P,S] :=  H_{Q_1C} + H_{Q_2C} = g_1\int dq_1dq_2dx\, P (\partial_{q_1} S) x +  g_2\int dq_1dq_2dx\, P (\partial_x  S) q_2,
\eeq
followed by a measurement of the classical position. Here  $g_1$ and $g_2$ are coupling constants, and $H_{Q_1C}$ and $H_{Q_2C}$ represent interactions between $Q_1$ and $C$ and between $Q_2$ and $C$, respectively.  It is straightforward to solve the corresponding Hamilton equations of motion~(\ref{motion}) to give
\begin{align} \label{pevolution}
P_t(q_1,q_2,x) &= P_0(q_1-g_1tx +\half g_1g_2t^2 q_2, q_2, x-g_2t q_2), \\
\label{sevolution}
S_t(q_1,q_2,x) &= S_0(q_1-g_1tx +\half g_1g_2t^2 q_2, q_2, x-g_2t q_2) ,
\end{align}
as per equations~(16) and~(17) of~\cite{HR2017}. If the measurement of the classical position is made at time $t$, with result $x=a$, then the marginal probability density $P_t(x)$ reduces to $P_t(x)=\delta(x-a)$, and hence the density operator $\rho_{QQ'}$ in equation~(\ref{equiv})  reduces to the form $|\psi_a\rangle\langle\psi_a|$. It follows, assuming that  the classical and quantum components are initially independent, with initial wave functions $\psi_1(q_1)$ and $\psi_2(q_2)$ for $Q_1$ and $Q_2$ and an initial ensemble, $(P_0(x),S_0(x))$ for $C$, that the post-measurement state of the quantum particles $Q_1$ and $Q_2$ at time $t$ is described by the wave function
\beq \label{psicond}
\psi_{t}(q_1,q_2|a) =K_a \psi_1(q_1-g_1ta +\half g_1g_2t^2 q_2)\, \psi_2(q_2)\sqrt{P_0(a-g_2tq_2)}e^{iS_0(a-g_2tq_2)/\hbar} , 
\eeq
 as per equation~(21) of~\cite{HR2017}, where $K_a$ is a normalisation constant. 
Clearly this does not factorise into functions of $q_1$ and $q_2$ in general, i.e., the quantum particles typically become entangled following the interaction and measurement.

\subsection{Classicality of the model}
\label{sec:classicality}
	
Marconato and Marletto claim that the classical particle $C$ in the above example is in fact not a `classical' system at all. However this claim relies on a number of errors and misunderstandings, which we explicitly address here. 

First, Marconato and Marletto assume that the Hamiltonian in equation~(\ref{ham}) corresponds to first interacting $Q_1$ with $C$ and then interacting $Q_2$ with $C$, whereas it actually describes a simultaneous interaction. This is not a fundamental difficulty, however, as if one instead applies $H_{Q_1C}$ and then $H_{Q_2C}$ (or, equivalently, takes $g_2=0$ and then $g_1=0$), each for a time $t$, this simply replaces the factor $\half g_1g_2$ appearing in equations~(\ref{pevolution})--(\ref{psicond}) by $g_1g_2$, which does not affect the entanglement properties of $\psi_t(q_1,q_2|a)$. Hence we may proceed by assuming this replacement has been made.

Second, Marconato and Marletto ignore the important step of conditioning on a measurement of the classical position, leading them to incorrectly assume that the  entanglement between $Q_1$ and $Q_2$ is described by equations~(\ref{pevolution}) and~(\ref{sevolution}) rather than by the wave function in equation~(\ref{psicond})---even though it was explicitly noted in~\cite{HR2017} that without such conditioning the particles  $Q_1$ and $Q_2$ are described by the density operator in equation~(22) thereof ({\it cf.} $\hat\rho_{Q_1Q_2}$ in~(\ref{equiv})), which need not be entangled. 

Third, Marconato and Marletto make a fundamental error in interpreting the fact that the ensemble {\it before} measurement, as described by equations~(\ref{pevolution}) and~(\ref{sevolution}), can be written in the quantum-like form
\beq \label{hybrid}
\psi_t(q_1,q_2,x) = e^{it(g_1 \hat p_1\hat x+g_2 \hat q_2\hat k)/\hbar}\, \psi_0(q_1,q_2,x) ,
\eeq
as per equation~(18) of our paper~\cite{HR2017}. Here $\psi_t(q_1,q_2,x)$ denotes the `hybrid wave function' $\sqrt{P_t}e^{iS_t/\hbar}$ and $\hat p_1$ and $\hat k$ denote the linear operators $(\hbar/i)\partial_{q_1}$ and $(\hbar/i)\partial_{x}$ conjugate to $\hat q_1\equiv q_1$ and $\hat x\equiv x$ (note, in relation to the first point above, that the exponential operator in~(\ref{hybrid}) is replaced by $e^{itg_2 \hat q_2\hat k/\hbar} e^{itg_1 \hat p_1\hat x/\hbar}$ for sequential interactions). In particular, Marconato and Marletto argue that that since $\hat k$ formally appears in the evolution as per equation~(\ref{hybrid}) (which observably changes the density operator describing $Q_1$ and $Q_2$, from $\hat\rho_{Q_1Q_2}(0)$ to $\hat\rho_{Q_1Q_2}(t)$), then (i)~$\hat k$ must represent a measurable observable of the classical particle $C$, and (ii)~hence $C$ cannot be truly classical because $[\hat x,\hat k]\neq 0$.

 However, this argument is logically flawed. \blk	To see this, it is useful to consider  a \blk related fully classical model. obtained by replacing the quantum particles $Q_1$ and $Q_2$ with classical particles $C_1$, $C_2$ having coordinates $x_1$ and $x_2$. Thus in this case we have two classical particles interacting via a classical mediator and there are no quantum particles present.  The Hamiltonian in equation~(\ref{ham}) then becomes
\begin{align}\label{hamc}
H'[P,S] &:=  H_{C_1C} + H_{C_2C} = g_1\int dx_1dx_2dx\, P (\partial_{x_1} S) x +  g_2\int dq_1dq_2dx\, P (\partial_x  S) x_2 \nn\\ 
&= C_{g_1 k_1x+g_2 kx_2},
\end{align}
where the last line follows from equation~(\ref{cf}), i.e., $H'[P,S]$ corresponds to the classical phase space Hamiltonian $h:=g_1 k_1x+g_2kx_2$. The equations of motion~(\ref{motion}) under $H'[P,S]$ are equivalent to the classical continuity equation for $P$ and the classical Hamilton-Jacobi equation for $S$ for Hamiltonian $h$ and, importantly, their solution is precisely as per equations~(\ref{pevolution}) and~(\ref{sevolution}), with $q_1$ and $q_2$ replaced by $x_1$ and $x_2$. Hence, this {\it fully classical} evolution  can be written in the quantum-like form
\beq \label{hybridc}
\psi'_t(x_1,x_2,x) = e^{it(g_1 \hat k_1\hat x+g_2 \hat x_2\hat k)/\hbar}\, \psi'_0(q_1,q_2,x) ,
\eeq
analogous to equation~(\ref{hybrid}), where $\psi_t'(x_1,x_2,x)$ denotes the `classical wave function' $\sqrt{P_t}e^{iS/\hbar}$ and $\hat k_1\equiv (\hbar/i)\partial_{x_1}$, $\hat k\equiv(\hbar/i)\partial_{x}$. But it is clearly  logically \blk absurd to argue, for this purely classical system,  that simply because $\hat k$ formally appears in the evolution as per equation~(\ref{hybridc}) (which observably changes the classical phase space density describing $C_1$ and $C_2$), then (i)~$\hat k$ must represent a measurable observable of $C$, and (ii) hence $C$ cannot be truly classical because $[\hat x,\hat k]\neq0$.  And it is similarly meaningless to argue the identical conclusion from equation~(\ref{hybrid}), as done by Marconato and Marletto~\cite{MM}. It is a logical error that confuses form  ($\hat k$) \blk with substance  (classical vs quantum). \blk

 Indeed, \blk the similar evolutions in equations~(\ref{hybrid}) and~(\ref{hybridc}) in fact arise from the  similar forms of the \blk bracket relations
\beq \label{posmom}
\{C_x,C_k\} = C_1 = 1, \qquad \qquad  \{Q_{\hat q},Q_{\hat p}\} = Q_{\hat 1} = 1 ,
\eeq
for classical and quantum position and momentum observables, following from equation~(\ref{brackets}).  These relations simply reflect the well-known \blk fact that momentum is the generator of translations, in {both classical and quantum mechanics. 

 Fourth,  \blk contrary to what is suggested by Marconato and Marletto,  the \blk `non-commutation' of classical position and momentum in Eq.~(\ref{posmom}) no more implies that particle $C$ satisfies their stated criterion for `nonclassicality' (i.e., that $C$ has {\it ``at least two variables that are necessary to describe its features, and yet cannot be measured to arbitrarily high accuracy simultaneously''}~\cite{MM}), than does the equivalent Poisson bracket relation $\{x,k\}=1$ on classical phase space.  Indeed, one can always couple the classical particle $C$ to two classical `pointer' particles $C_1$ and $C_2$, irrespective of the presence of $Q_1$ and $Q_2$, via a suitable phase space Hamiltonian $h$ and corresponding ensemble Hamiltonian $C_h$, to jointly measure its position and momentum to an arbitrary accuracy. This is a simple consequence of classical phase space dynamics (which places no constraints on joint measurement accuracy) and the one--one correspondence between phase space functions and classical observables in equation~(\ref{brackets}), with the measured joint probability density in the ideal limit given by $\rho_C(x,k)$ in equation~(\ref{equiv}). We comment further on definitions of classicality and nonclassicality in section~\ref{sec:additional}).

Fifth and finally, Marconato and Marletto claim that there is an `inherent ambiguity' in how to calculate the statistics of the quantum particles $Q_1$ and $Q_2$ following the interaction in our example (e.g., to check whether they are entangled). This is clearly wrong, however, as we explicitly give the density operator for calculating these statistics in equation~(22) of~\cite{HR2017}, equivalent to $\hat\rho_{Q_1Q_2}$ in Eq.~(\ref{equiv}) above.  For example, the expectation value of the product of two quantum operators  $\hat M_1$ and $\hat M_2$, acting on the Hilbert spaces of $Q_1$ and $Q_2$ respectively, follows from either of equations~(\ref{qm}) or~(\ref{equiv}) as
\beq
\langle \hat M_1\otimes \hat M_2\rangle = \tr{\hat \rho_{Q_1Q_2} \hat M_1\otimes \hat M_2} ,
\eeq
allowing the calculation of the values of entanglement witnesses, etc., and is measured in the usual way by coupling the quantum particles to suitable measuring apparatuses.

\subsection{Locality of the model}
\label{sec:locality}	
	
Marconato and Marletto  also \blk claim that our model {\it ``violates the principle of no action at a distance''}~\cite{MM}, and hence that it is nonlocal.  However, this claim is very easily demonstrated to be incorrect, not only for the specific model criticised by Marconato and Marletto, but far more generally. \blk

%this is so clearly not the case that it is simple to explicitly not so when the interactions $H_{Q_1C}$ and $H_{Q_2C}$ between the quantum and classical particles are applied sequentially (as assumed by Marconato and Marletto and discussed in the first point of section~\ref{sec:classicality} above). In particular, 

 First, \blk under the action of $H_{Q_1C}$ in equation~(\ref{ham}) the average value of any observable $Q_{\hat M_2}$ of $Q_2$ evolves as per equation~(\ref{dvdt}), i.e.
\beq \label{local1}
\frac{d\langle \hat M_2\rangle}{dt} =\{H_{Q_1C}, Q_{\hat M_2}\} =0,
\eeq
where the final equality follows on substituting the forms of the observables from equations~(\ref{qm}) and~(\ref{ham}).  Thus the interaction between $C$ and $Q_1$ has no effect on any observable of $Q_2$, i.e., the interaction is local.  Similarly, under the action of $H_{Q_2C}$, 
\beq \label{local2}
\frac{d\langle \hat M_1\rangle}{dt} =\{H_{Q_2C}, Q_{\hat M_1}\} =0
\eeq	
for any operator $\hat M_1$ of particle $Q_1$, i.e., the interaction between $C$ and $Q_2$ is also local.	

Marconato and Marletto  indeed \blk acknowledge that the model is local in the above sense, but assert this is a ``fortunate coincidence'' that does not hold for all possible choices of ensemble Hamiltonian. This is logically irrelevant, however, in that the particular model under discussion {\it is} local in the above sense: one could as well assert that classical dynamics is nonlocal simply because there are formal choices of  classical \blk Hamiltonians that describe nonlocal interactions.
	
 Second, \blk Marconato and Marletto note that  the configuration-ensemble formalism more generally need \blk  not satisfy the principle of 
 `strong separability'~\cite{PRA2008, HRbook}, i.e., there are classical and quantum observables for which $\{C_f,Q_{\hat M}\}\neq 0$. This is indeed the case, as is extensively discussed in~\cite{HRbook}. However,  this property is only relevant to the locality of the model if such observables appear in the ensemble Hamiltonian---and they do not  for the ensemble Hamiltonian in Eq.~(\ref{ham}), as evidenced by Eqs.~(\ref{local1}) and~(\ref{local2}) above. \blk Locality properties of Bohmian and mean-field models, referred to by Marconato and Marletto, are similarly irrelevant to our model.
 
  Further, the property of strong separability does in fact hold whenever the classical system is initially independent from the quantum systems~\cite{PRA2008, HRbook}. Explicitly, if the initial ensemble $(P_0,S_0)$ is of the form
 \beq
 P_0(q,x)=P_Q(q)P_C(x),\qquad S_0(q,x) = S_Q(q) +S_C(x),
 \eeq
 then the ensemble satisfies
 \beq
 \{ C_f,Q_{\hat M} \} = 0 
 \eeq
 at all times $t\geq0$~\cite{PRA2008, HRbook}. Such initial independence is in fact a required assumption of the no-go theorem quoted by Marconato and Marletto~\cite{MM} (see also section~\ref{sec:example} below).  Hence, any configuration-ensemble model relevant to this theorem is in fact {\it fully local} in the sense of strong separability. \blk

Finally, we note for interest (as has been pointed out elsewhere~\cite{HRgrav,HRbook}), that for the case of quantum field entanglement generated via general relativistic interactions with classical spacetime, the issue of strong separability is moot: the fields are never gravitationally decoupled from the spacetime, due to the nature of the interaction, so that there is always nonlocality in this sense. A similar effect can be seen even in the semiclassical limit of two quantum systems separated by a fixed distance $d$ with Hamiltonians $\hat h_1$ and $\hat h_2$, for which the joint Hamiltonian has the nonlocal form~\cite{Ruiz}
\beq
\hat h_{12} = \hat h_1\otimes \hat 1 + \hat 1\otimes \hat h_2 - \frac{G}{c^4d} \hat h_1\otimes\hat h_2 .
\eeq
In particular, unlike equations~(\ref{local1}) and~(\ref{local2}), as long as there is a gravitational interaction (i.e., $\hat h_1$ and $\hat h_2$ are not constant), then there are always observables of each system that are influenced by this interaction.

	\section{What the no-go theorems for entanglement generation actually rule out}
	\label{sec:assump}
	
	\subsection{A simple example}
	\label{sec:example}
	
	Suppose that one of a pair of two-qubit states, $\hat\rho_0$ or $\hat\rho_1$, is shared between observers $A$ and $B$, according to the value of a classical bit $c=0$ or~$1$.  Thus, if the values of $c$ are equally likely, the qubit measurement statistics are described by the density operator $\hat\rho=\half(\hat\rho_0+\hat\rho_1)$. For the special case that $\hat\rho_0$ and $\hat\rho_1$ are two orthogonal Bell states, the qubits are unentangled (e.g., for the choice $\hat\rho_c=\frac14(\hat1\otimes\hat1-\hat X\otimes \hat X+(-1)^c \hat Y\otimes \hat Y+(-1)^c \hat Z\otimes\hat Z)$, where $\hat X,\hat Y,\hat Z$ are the Pauli spin observables, one has the separable decomposition $\hat \rho=\half(\frac{1+\hat X}{2}\otimes\frac{1-\hat X}{2}+ \frac{1-\hat X}{2}\otimes\frac{1+\hat X}{2}$).  Further, for this special case there is always a local unitary transformation $\hat U_A$ on observer $A$'s qubit such that $\hat U_A\otimes \hat 1 \hat \rho_1\hat U_A^\dagger\otimes \hat 1=\hat \rho_0$ (e.g., $\hat U_A=\hat X$ for the above choice of $\hat \rho_c$). Hence, if the classical bit $c$ is communicated to observer $A$, and she performs the local unitary transformation $(\hat U_A)^c$, the qubit statistics are subsequently described by the density operator $\hat \rho'=\hat \rho_0$, i.e., the qubits are now maximally entangled.
	
	The above example is a simplification of a related example by Krisnanda {\it et al.}~\cite{Kris2017}, and shows that entanglement can be generated between two qubits via an interaction mediated by a classical bit. Further, this interaction, described by $(\hat U_A)^c$, is clearly local, i.e., the principle of locality is satisfied. The principle of ``interoperability of information'' referred to by Marconato and Marletto~\cite{MM} is also trivially satisfied (i.e., noting that the classical bit can be formally encoded in a qubit basis, information in distinguishable states can be permuted and copied).  It follows that any no-go theorem ruling out such entanglement generation must rely on an assumption that goes {\it beyond} these two principles. This invalidates the general claim made by Marconato and Marletto that {\it ``when observing entanglement \dots one can rule out all classical theories of gravity obeying the above-mentioned general principles''}~\cite{MM}.
	
	Entanglement generation in the above example is possible because the classical bit contains information about the preparation of the ensemble. Moreover, it is in fact physically reasonable for a classical gravitational field to carry such information, if the preparation devices for $\hat \rho_0$ and $\hat \rho_1$ are gravitationally distinguishable.
	So, how do the various no-go theorems in the literature avoid this possibility? By making formal assumptions \textit{that rule it out from the start}. 
	
	For example, for no-go results based on modelling quantum-classical interactions by local operations and classical communication (LOCC) on a quantum state $\hat \rho$~\cite{Kafri,Bose}, the formal definition of LOCC precludes any dependence on how  $\hat \rho$ is prepared, and hence rules out encoding such preparation information in a classical bit.  For no-go results based on Koopman-type models of quantum-classical interactions~\cite{MVprl, MV2020}, corresponding to representing the classical components and their evolution by operators diagonal in some `classical basis' or by `maximum information observables', it is assumed that the quantum components are initially uncorrelated with the classical components, thus ruling out any initial correlation with a classical bit (if an initial correlation {\it is} allowed in such models, then the generated entanglement is bounded by the corresponding mutual information~\cite{Kris2020}). Finally, a no-go result for mean-field models of quantum-classical interaction~\cite{HR2017} similarly assumes that the classical phase space observables are initially uncorrelated with the quantum components.
	
	The above example demonstrates the logical point that the various no-go theorems in the literature rely on assumptions that go beyond the principles of locality and interoperability of information, and in particular invalidates the above-quoted general claim by Marconato and Marletto. The same logical point also applies to our (rather different) models of entanglement generation in~\cite{HR2017}, as we next demonstrate.

	\subsection{Additional assumptions made in the no-go theorem of Ref.~\cite{MV2020}}
	\label{sec:additional}
	
	Similarly to the above example,  our configuration-ensemble models for the generation of entanglement between two quantum systems $Q_1$ and $Q_2$ via a classical mediator $C$ are compatible with the no-go theorem in Ref.~\cite{MV2020}. In particular, contrary to the general claim by Marconato and Marletto, the proof of this theorem explicitly requires a number of additional assumptions over and above the principles of locality and interoperability of information, that are not satisfied by our models in~\cite{HR2017}. These assumptions include, for example~\cite{MV2020}:
	\begin{enumerate}
		
		\item the quantum systems $Q_1$ and $Q_2$ are qubits (whereas they are quantum particles with continuous degrees of freedom in the model criticised by Marconato and Marletto---see also section~\ref{sec:model} above);
		
		\item the classical system $C$ is described by a `binary maximum information observable', i.e., by a bit (whereas it is a classical particle with continuous degrees of freedom in our model);
		
		\item the interactions between $Q_1$ and $C$ and between $Q_2$ and $C$ have identical forms (whereas they have different but closely-related forms in our model, as per~(\ref{ham}));
		
		\item these interactions have the property of evolving two particular uncorrelated initial states of $Q_1$ and $Q_2$ to two orthogonal maximally entangled two-qubit states (which is not a property of our model);
		
		\item entanglement is generated without any measurement on the classical system (whereas in our model the entanglement is conditional on making a measurement on the classical system).
		
	\end{enumerate}
	The presence of all these additional assumptions implies that the no-go theorem is simply not applicable to our models in~\cite{HR2017}. Thus these models present no challenge to the mathematical validity of the theorem, and there is no {\it a priori} need for its `vindication' as sought by Marconato and Marletto. 
	
	Finally, it is worth noting that there is a further significant difference between our models and the theorem in Ref.~\cite{MV2020}, concerning the very definition of `classical' and `nonclassical'.  In particular, \textit{in our definition of classicality the observables of a classical system are in one--one correspondence with functions of position and momentum on a classical phase space}: these observables obey the classical Poisson bracket algebra of such functions at all times (see also section~\ref{sec:model}). A similar notion of classicality is used in mean-field models of quantum-classical interaction~\cite{HR2017,mfpapers,mfpapers2}, but not in models in which  classical observables are described by LOCC~\cite{Kafri,Bose} or `commute'~\cite{MVprl,Kris2017, Kris2020,koopmanpapers,koopmanpapers2}.  
	
	In contrast Ref.~\cite{MV2020} does not directly define `classical', but instead gives a criterion for a system to be `nonclassical'.  The informal statement of this criterion, that nonclassical systems have incompatible observables, has already been considered in section~\ref{sec:classicality} and shown not to apply to our models.  However, the formal statement of the criterion, on page~4 of~\cite{MV2020}, is much stronger (and rather non-intuitive). In particular, it requires a nonclassical system be able to {\it ``enable non-classical tasks on other superinformation media, such as establishing entanglement''}~\cite{MV2020}, where quantum systems are particular examples of `superinformation media'.  This suggests that the theorem in Ref.~\cite{MV2020}, showing that entanglement generation can only be enabled via a mediator that is `nonclassical' in the above sense, has a somewhat circular flavour in comparison to other no-go theorems~\cite{Kafri,Bose,MVprl,HR2017}.

	\section{Conclusion}
	As well as pointing out the errors and misconceptions underlying the claims made by Marconato and Marletto, this Comment has provided a welcome opportunity for us to clarify our models of entanglement generation via a classical mediator, and to examine how such models evade the various no-go theorems in the literature.  Furthermore, as the simple example discussed in section \ref{sec:assump} demonstrates, these no-go theorems rely on assumptions that go beyond the principles of locality and interoperability of information. Thus we have to conclude that the general claim made by Marconato and Marletto, that {\it ``when observing entanglement \dots one can rule out all classical theories of gravity obeying the above-mentioned general principles''}~\cite{MM}, is simply incorrect.

	{\flushleft {\it Acknowledgements:} MH is grateful to Tomasz Paterek for helpful discussions in 2019.}\\

\end{document}